\documentclass[nocopyright]{cimento}
\usepackage{physics}
\usepackage{tikz}
\usepackage{slashed}
\usepackage{pgfplots}
\usepackage{mathtools}
\usepackage{siunitx}
\usepackage{physics}
\usepackage[compat=1.1.0]{tikz-feynman}
\usepackage[T1]{fontenc}
\usepackage{subcaption}
\usepackage{tabularx}
\usepackage{multirow}
\usepackage{booktabs}
\usepackage{cleveref}
\usepackage{tabularray}
\usepackage{nicematrix}
\usepackage{adjustbox}
\usepackage{siunitx}

\usepackage{unicode-math}

\newcommand{\sw}{s_{W}} 
\newcommand{\cw}{c_{W}} 

\definecolor{mypurple}{rgb}{0.31, 0.08, 0.41} \definecolor{myblue}{rgb}{0.67, 0.25
, 0.33
} \definecolor{mygreen}{rgb}{0.63, 0.988, 0.73} \definecolor{myyellow}{rgb}{0.92, 0.55, 0.22} %

\title{On the new physics in Bhabha luminometry \\at future $e^+e^-$ colliders}

\author{Clara L. Del Pio\from{BNL} \atque Francesco P. Ucci \from{unipv}\from{infn-pv}\thanks{Speaker.}} 
\instlist{
\inst{BNL}{Department of Physics, Brookhaven National Laboratory,
Upton, New York 11973, USA}
\inst{unipv}{Dipartimento di Fisica "Alessandro Volta", Universit\`a di Pavia, Via A. Bassi 6, 27100 Pavia, Italy}
\inst{infn-pv}{INFN, Sezione di Pavia, Via A. Bassi 6, 27100 Pavia, Italy}}
\begin{document}
\maketitle
\begin{abstract}
The absolute machine luminosity is a key quantity to achieve the high-precision physics program of future $e^+e^-$ collider. It is determined by measuring a theoretically well-known process, which, ideally, can be computed with arbitrary precision in the perturbation theory. However, yet undiscovered new physics could give a non-negligible contribution to the cross section of the luminosity monitoring process, thus invalidating the uncertainty determination of measured quantities. We assess the theoretical error of non-Standard Model origin to the small-angle Bhabha scattering in various future colliders scenarios. In addition, a possible running strategy to constrain unknown heavy interactions is proposed, relying on asymmetries that do not depend on the absolute luminosity.
\end{abstract}
\vspace{-1.4cm}
\section{Introduction}
\begin{figure}[t]
\begin{equation*}
    \begin{gathered}
    \vspace{0.1cm}
\begin{tikzpicture}
  \begin{feynman}[small]
            \vertex (a);
            \vertex[below=1.25cm of a] (b);
    \vertex[above left=0.375cm and 1cm of a] (c) ;
    \vertex[below left=0.375cm and 1cm of b] (d) ;
    \vertex[above right=0.375cm and 1cm of a] (e);
    \vertex[below right=0.375cm and 1cm of b] (f);
    \diagram* {
      (a) -- [photon,edge label=$\gamma$] (b),
      (d) -- [fermion] (b),
      (a) -- [fermion] (c),
      (e) -- [fermion] (a),
      (b) --[fermion] (f),
    };
  \end{feynman}
\end{tikzpicture}
    \end{gathered} \qquad\quad
    \begin{gathered}
    \vspace{-0.5cm}
\begin{tikzpicture}
  \begin{feynman}[small]
            \vertex (a);
            \vertex[below=1.25cm of a ,style=dot, fill= mypurple, draw=mypurple, minimum size=8pt, label={[yshift=-6pt]below:$\Delta g_{L/R}^{Ze}$}] (b) {};
    \vertex[above left=0.375cm and 1cm of a] (c) ;
    \vertex[below left=0.375cm and 1cm of b] (d) ;
    \vertex[above right=0.375cm and 1cm of a] (e);
    \vertex[below right=0.375cm and 1cm of b] (f);
    \diagram* {
      (a) -- [photon,edge label=$Z$] (b),
      (d) -- [fermion] (b),
      (a) -- [fermion] (c),
      (e) -- [fermion] (a),
      (b) --[fermion] (f),
    };
  \end{feynman}
\end{tikzpicture}
    \end{gathered} 
    \quad\quad
   \begin{gathered}
\begin{tikzpicture}
  \begin{feynman}[small]
    \vertex[style=dot, minimum size=8pt,fill=myblue, draw=myblue, label={[yshift=-7pt]below:$C_{i}$}] (a) {};
    \vertex[above left=1cm and 1cm of a] (c) ;
    \vertex[below left=1cm and 1cm of a] (d) ;
    \vertex[above right=1cm and 1cm of a] (e);
    \vertex[below right=1cm and 1cm of a] (f);
    \diagram* {
      (d) -- [fermion] (a),
      (a) -- [fermion] (c),
      (e) -- [fermion] (a),
      (a) --[fermion] (f),
    };
  \end{feynman}
\end{tikzpicture}
    \end{gathered}\quad \qquad \begin{gathered}
\begin{tikzpicture}
  \begin{feynman}[small]
    \vertex[style=dot, minimum size=8pt,fill=myblue, draw=myblue, label={[yshift=-7pt]below:$C_{i}$}] (a) {};
    \vertex[above left=1cm and 1cm of a] (c) ;
    \vertex[below left=1cm and 1cm of a] (d) ;
    \vertex[above right=1cm and 1cm of a] (e);
    \vertex[below right=1cm and 1cm of a] (f);
    \vertex[above left=0.75cm and 0.75cm of a] (g);
    \vertex[above right=0.75cm and 0.75cm of a] (h);
    \diagram* {
      (d) -- [fermion] (a),
      (a) -- [fermion] (c),
      (e) -- [fermion] (a),
      (a) --[fermion] (f),
      (g) --[photon,looseness=0.5] (h),
    };
  \end{feynman}
\end{tikzpicture}
    \end{gathered}
    \qquad
    \begin{gathered}
    \vspace{-10pt}
    \begin{tikzpicture}
  \begin{feynman}[small]
            \vertex (a);
            \vertex[below=1.25cm of a ,style=dot, fill= myyellow, draw=myyellow,label={[yshift=-3pt]below:$C_{j}$}, minimum size=8pt] (b) {};
    \vertex[above left=0.375cm and 1cm of a] (c) ;
    \vertex[below left=0.375cm and 1cm of b] (d) ;
    \vertex[above right=0.375cm and 1cm of a] (e);
    \vertex[below right=0.375cm and 1cm of b] (f);
    \vertex[below=0.75 cm of a](g);
    \diagram* {
      (a) -- [photon,edge label=$\gamma\text{,}Z$] (g),
      (d) -- [fermion] (b),
      (a) -- [fermion] (c),
      (e) -- [fermion] (a),
      (b) --[fermion] (f),
      (g) --[fermion,half left] (b) --[fermion, half left] (g),
    };
  \end{feynman}
\end{tikzpicture}
\end{gathered}
\end{equation*}
\caption{Diagrams contributing to the SABS cross section. The first diagram is the SM $t$-channel photon exchange. The second two are LO SMEFT diagrams representing a modified $Zee$ vertex and a four-fermion interaction. The latter two are next-to-leading-order (NLO) diagrams in the SMEFT.}
\label{fig:HNPdiags}
\end{figure}
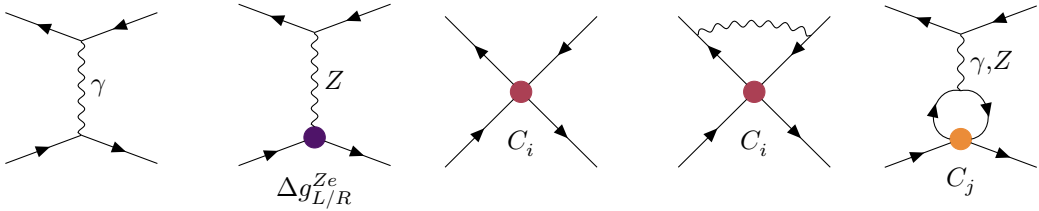

At $e^+e^-$ colliders, the measurement of the absolute luminosity $L$ is a source of systematic uncertainty that enters every absolute cross section determination, and it is performed by relying on a reference process
\begin{equation}
    L = \int \mathcal{L}\,\dd t=\frac{N_0}{\epsilon\sigma_0}\,.
\end{equation}
Here, $N_0$ is the number of events for such process, $\sigma_0$ its theoretical cross section, while $\epsilon$ is the experimental acceptance. Therefore, in order to minimise the relative error $\delta L$, such process has to have a large cross section, a clean experimental signature, as well as being calculable at high orders in the perturbation theory. At LEP, such requests were met by the small angle Bhabha scattering (SABS)~\cite{LEP_SABS}, being essentially a pure QED process dominated by the photon exchange (shown in the first diagram of Figure~\ref{fig:HNPdiags}), with the $Z$ contribution being a small leading-order (LO) correction. 
At future Higgs, top and electroweak factories~\cite{ECFA}, the precision goal will be even more challenging, seeking for $10^{-4}$ precision or better at the $Z$ pole runs and at the $WW$ production threshold, while being at the order of $10^{-3}$ for energies larger than the $ZH$ threshold. If the uncertainty on the luminosity is kept under the $10^{-4}$ level, for instance, the measurement of the mass and width of the $W$ boson will be allowed with $\order{1~\rm{MeV}}$ uncertainty~\cite{Wmass}.\\
While the path for improving SM predictions is drawn clearly in the literature~\cite{path}, the impact of an unknown degree of freedom (d.o.f.) mediating the SABS is still not assessed. We address the question of quantifying the level of theoretical uncertainty caused by the unconstrained parameter space of new physics (NP) for various $e^+e^-$ colliders proposals~\cite{lumi}. After the beyond the Standard Model (BSM) effect is estimated, we propose a strategy to overcome such uncertainties and secure the high precision program of future machines.
\section{New physics in SABS}
\begin{figure}[h]
    \centering
    \includegraphics[width=0.85\linewidth]{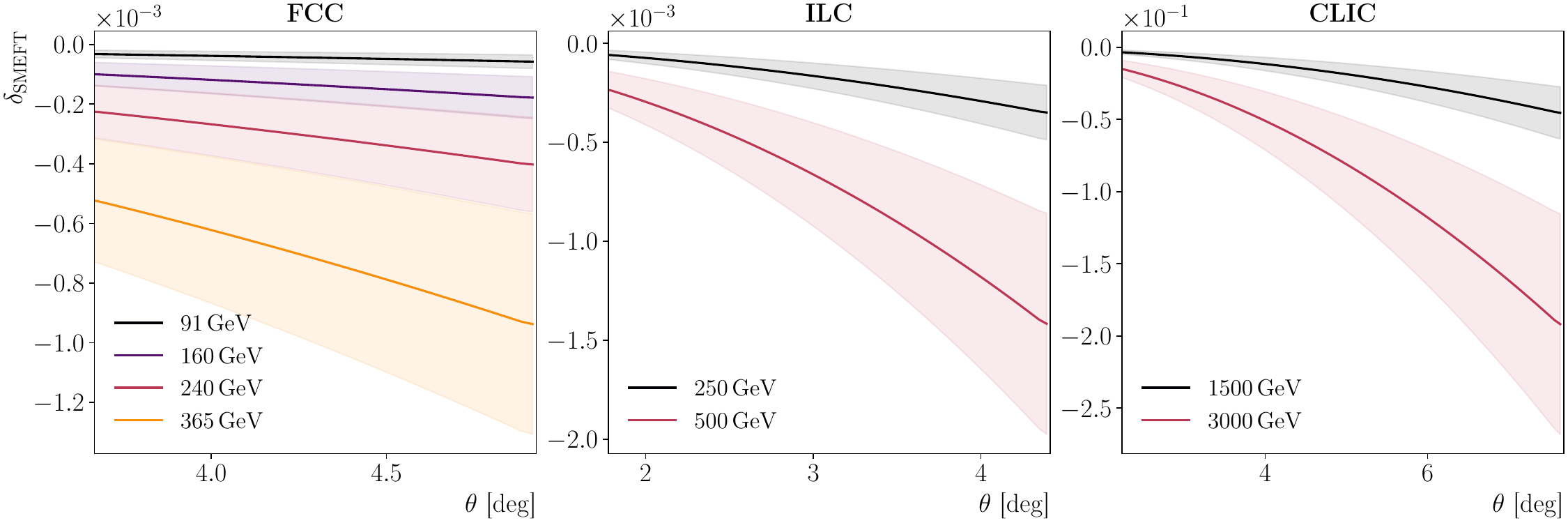}
    \caption{Differential deviation of the SABS cross section as computed in the SMEFT, according to Eq.~\eqref{eq:deltasmeft}.}
    \label{fig:SMEFT_diff}
\end{figure}

The SABS cross section can be contaminated by light mediators with feeble couplings to the SM particles or by heavy new d.o.f. In the case of light new physics (LNP) one may specify the spin and the parity of the new particles, therefore introducing a certain model. In the latter case, heavy new physics (HNP) can be parametrised by means of the Standard model effective field theory (SMEFT), that considers the SM as the low-energy approximation of some UV theory in a model agnostic way.\\
In this analysis, the most promising future colliders projects are taken into account, namely FCC~\cite{fcc}, CEPC~\cite{cepc}, ILC~\cite{ilc} and CLIC~\cite{clic}, whose details are reported in Table~\ref{tab:SMEFT_numbers}. We adopt the $\{\alpha(M_Z),G_\mu,M_Z\}=\{1/127.95,1.1668\times10^{-5}~{\rm GeV}^{-2},M_Z=91.1876~{\rm GeV}\}$ input scheme for the EW sector. The relevant amplitudes have been calculated using \textsc{FeynCalc} and \textsc{FeynArts}~\cite{feynarts}, with the \textsc{SmeftFR}~\cite{smeftfr} model. The numerical results have been produced using an updated version of \textsc{BabaYaga@NLO}~\cite{babayaga} Monte Carlo (MC) generator.\\
In Ref.~\cite{lumi}, it was shown that the LNP contribution is negligible, being at most of the order of $\delta_\text{LNP}\sim6\times 10^{-6}$ for the hypothetical $X_{17}$ particle, whose existence is still under scrutiny. Therefore, here we focus on the hypothesis of the new physics scale lying far above the EW scale $\Lambda_\text{NP}\gg \Lambda_\text{EW}$. In this scenario, the SM Lagrangian can be extended with all possible higher-dimensional operators, built upon the same fields and symmetries of the SM and suppressed by powers of $\Lambda_\text{NP}$
\begin{equation}
    \mathcal{L}_\text{SMEFT} = \mathcal{L}_\text{SM} + \sum_i \frac{C_i \hat O_i^{(6)}}{\Lambda^2_\text{NP}} + \order{\frac{1}{\Lambda^4_\text{NP}}}\, ,
\end{equation}
where the expansion is truncated at dimension-six in order to capture the leading effects.  The EW Lagrangian is therefore modified as~\cite{falkowski}
\begin{equation}
\begin{alignedat}{2}
\mathcal{L}^{\text{EW}}_{\text{SMEFT}} = &-\sqrt{4\pi\alpha} \,(
\bar{e} \gamma^\mu e) A_\mu \\
&
+ \frac{\sqrt{4\pi\alpha}}{\sw\cw}\Biggl[\bar{e}_L\gamma^\mu \left(\hat{g}_L+
\frac{\Delta g_L^{Ze}}{\Lambda^2_\text{NP}}\right) e_L
+\bar{e}_R \gamma^\mu \left(\hat{g}_R+ \frac{\Delta g_R^{Ze}}{\Lambda^2_\text{NP}}\right) e_R \Biggr] Z_\mu\, ,
\end{alignedat}
\label{eq:eight}
\end{equation}
with $A_\mu,Z_\mu$ being the photon and $Z$ boson fields, respectively, while $e_{L,R}$ are the left/right projections of the electron field. The SM left (right) couplings of the $Z$ boson to electrons $\hat g_L=s_W^2-1/2$ ($\hat g_R=s_W^2$) are shifted by the effective deviation $\Delta g^{Ze}_{L(R)}$ wich can be written as a linear combination of Wilson Coefficients (WCs) in the Warsaw basis. In addition, four-electron contact operator -- absent in the SM -- can contribute to the SABS cross section and are given by the Lagrangian 
\begin{equation}\label{EQ:4fermions}
    \begin{aligned}
\mathcal{L}_{\rm{SMEFT}}^{{4f}}=  \frac{1}{2}\frac{C_{ll}}{\Lambda^2_\text{NP}}&\left(\bar{e}_L \gamma^\mu e_L\right)\left(\bar{e}_L \gamma_\mu e_L\right) 
   +\frac{C_{le}}{\Lambda^2_\text{NP}}\left(\bar{e}_L \gamma^\mu e_L\right)\left(\bar{e}_R\gamma_\mu e_R\right)
   \\
   +\frac{1}{2}\frac{C_{ee}}{\Lambda^2_\text{NP}}&\left(\bar{e}_R \gamma^\mu e_R\right)\left(\bar{e}_R \gamma_\mu e_R\right)\, .
       \end{aligned}
\end{equation}
The effective theory is valid if the NP scale is at least two orders of magnitude larger than the scattering energy of the process, which is verified up to the CLIC setup, since the SABS $t$ channel probes energy as high as $100~\rm{GeV}$. The relevant diagrams contributing to the SABS are shown in Figure~\ref{fig:HNPdiags}.\\
In order to quantify the deviation with respect to the SM cross section, we define the relative differential shift as
\begin{equation}(\delta\pm\Delta\delta)_{\rm{SMEFT}}=\frac{1}{\sigma_{\rm{SM}}}\hspace{-1mm}\left( \sigma^{(6)}\pm \sqrt{\sum_{ij}\sigma^{(6)}_iV_{ij}\sigma_j^{(6)}}\right)\hspace{-1mm}, \qquad \sigma^{(6)}=\sum_{i=1}^n\frac{C_i}{\Lambda_\text{NP}^2}\sigma_i^{(6)}
\label{eq:deltasmeft}
\end{equation}
where the dimension-six cross section is given by the interference of the SM amplitude with the SMEFT one $\sigma_i^{(6)}= 2 \Re \mathcal{M}^\dagger_\text{SM}\mathcal{M}^{(6)}_{{\rm{SMEFT}},i}$ and the uncertainty is due to the covariance between WCs, with $V_{ij}=\Delta C_i\,\rho_{ij}\,\Delta C_j$. The numerical values and uncertainties for $C_i$ as well as their correlation matrix are given in Ref.~\cite{falkowski,lumi}. We focus on the four-fermion operators as the shifts of the $Zee$ couplings are well constrained and yield a negligible contribution.
\begin{table}[ht]
\centering
\caption{SMEFT contribution to SABS as defined in Eq.~\eqref{eq:deltasmeft} for various future $e^+ e^-$ facilities with the corresponding luminosity target precision $\Delta L/L$.}
\begin{tabular}{lcccr}
\toprule
Exp. & $[\theta_{\text{min}}, \theta_{\text{max}}]$ & $\sqrt{s}$ [GeV] & $(\delta \pm \Delta\delta)_{\rm{SMEFT}}$ & $\Delta L / L$ \\ 
\midrule
\multirow{4}{*}{FCC} & \multirow{4}{*}{$[\ang{3.7}, \ang{4.9}]$} & 91 & $(-4.2 \pm 1.7) \times 10^{-5}$ & $<10^{-4}$ \\
 &  & 160 & $(-1.3 \pm 0.5) \times 10^{-4}$ & \multirow{3}{*}{$10^{-4}$} \\
 &  & 240 & $(-2.9 \pm 1.2) \times 10^{-4}$ & \\
 &  & 365 & $(-6.7 \pm 2.7) \times 10^{-4}$ & \\
\midrule 
\multirow{2}{*}{ILC} & \multirow{2}{*}{$[\ang{1.7}, \ang{4.4}]$} & 250 & $(-1.2 \pm 0.5) \times 10^{-4}$ & \multirow{2}{*}{$<10^{-3}$} \\
 &  & 500 & $(-4.9 \pm 1.9) \times 10^{-4}$ & \\
\midrule
\multirow{2}{*}{CLIC} & \multirow{2}{*}{$[\ang{2.2}, \ang{7.7}]$} & 1500 & $(-9.7 \pm 3.9) \times 10^{-3}$ & \multirow{2}{*}{$<10^{-2}$} \\
 &  & 3000 & $(-4.2 \pm 1.7) \times 10^{-2}$ & \\
\bottomrule
\end{tabular}
\label{tab:SMEFT_numbers}
\end{table}  
 In Table~\ref{tab:SMEFT_numbers}, we give the numerical results for the future $e^+e^-$ colliders scenarios compared to the relative precision goal on luminosity. As one can see, the effects are non-negligible, especially for the $Z$ pole run at FCC-ee. Figure~\ref{fig:SMEFT_diff} shows the differential deviations in the luminometer acceptance for the same same setup, showing that, at small angles, the deviation is an increasing function of $\theta$.
\\
The NLO contribution in the SMEFT for the SABS cross section is given by the classes of diagrams depicted in the last two panels of Figure~\ref{fig:HNPdiags}. The former is a NLO EW correction to a dimension-six operator already appearing at tree-level, changing the LO SMEFT prediction at the $\order{1\%}$ scale. The latter class of contribution is a NLO insertion of a dimension-six operators, which is expected to be more relevant at higher energies; however, its contribution can be estimated to be $\delta_{\text{NLO},j}\sim 2\times10^{-4}C_j$. Therefore, the order of magnitude of the LO effects shown in Table~\ref{tab:SMEFT_numbers} due to the current bounds on four-fermions coefficients $\vec{C}_{4f}=(C_{ll},C_{le},C_{ee})$ is unchanged by NLO corrections and put serious limits on the high precision program of future colliders
\section{Asymmetries}
\begin{figure}
\begin{subfigure}{0.3\textwidth}
     \centering
    \includegraphics[width=\linewidth]{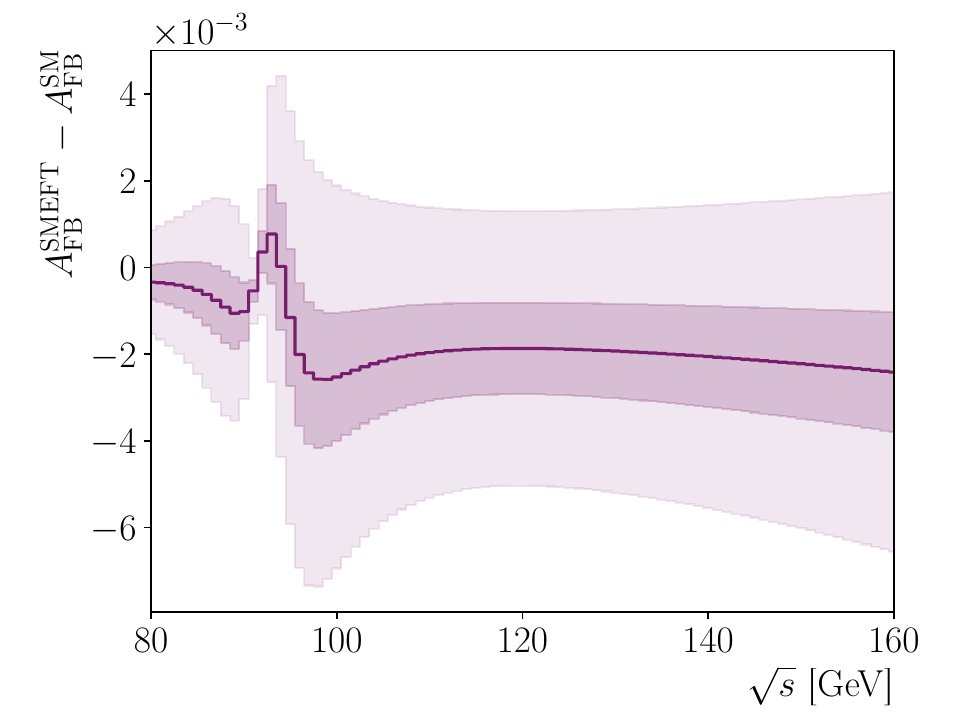}
\end{subfigure}
\begin{subfigure}{0.7\textwidth}
    \centering
    \includegraphics[width=\linewidth]{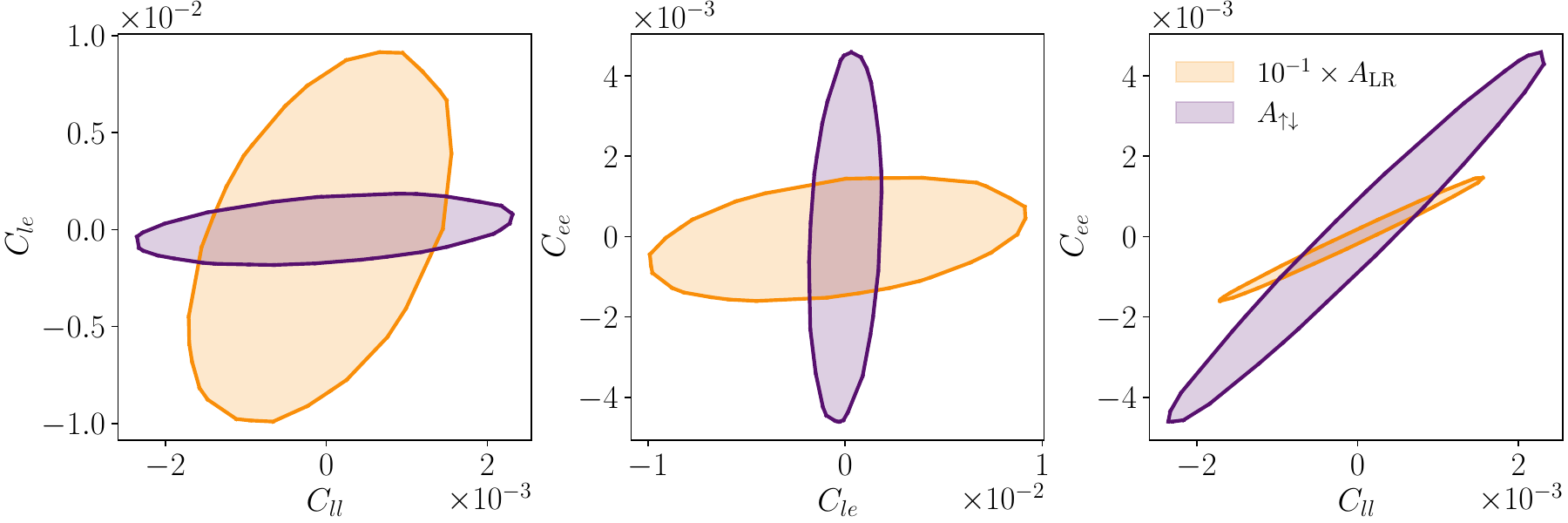}
\end{subfigure}
    \caption{The first panel shows the absolute difference between the SMEFT and SM $A_\text{FB}$ as a function of the energy. The latter three panels are the projection on $(C_i,C_j)$ planes of the $\chi^2\leq1$ region (Eq.\eqref{chi2}) for polarisation asymmetries.}
        \label{fig:ASY_pol}
\end{figure}

As such bounds will not change considerably by the HL-LHC data~\cite{celada} and by the start of future $e^+e^-$ colliders program, we explore the possibility of putting stronger bounds on $\vec{C}_{4f}$. In this scenario, luminosity-independent observables like asymmetries can be exploited to reduce the SMEFT uncertainty in the SABS. Writing the theoretical prediction for a generic 
asymmetry as follows
\begin{equation}\label{eq:ASYab}
    A_{ab}^\text{th}= A_{ab}^\text{SM}\left
\{1+\frac{(\sigma_{a}-\sigma_{b})^{(6)}}{(\sigma_{a}-\sigma_{{{b}}})_{\rm{SM}}}-\frac{(\sigma_{a}+\sigma_{b})^{(6)}}{(\sigma_{a}+\sigma_{b})_{\rm{SM}}}\right\}\, ,
\end{equation}
one can leverage the dependency on the $C_i$ given by $\sigma^{(6)}_{a,b}$ in order to fit the WCs to data. In particular, we consider asymmetries of the large angle Bhabha scattering (LABS), in order to explore an angular region complementary to the SABS, $i.e.$ $\theta\in[\ang{40},\ang{140}]$, and to enhance the $s$-channel contribution to the Bhabha. \\
For future colliders running at the $Z$ resonance, like FCC/CEPC, the natural asymmetry is the forward-backward $A_\text{FB}=(\sigma_\text{F}-\sigma_\text{B})/(\sigma_\text{F}+\sigma_\text{B})$, which is the difference of the integral of the cross section in the $\cos\theta>0$ ($\cos\theta<0$) region, normalised to the sum. By studying the absolute difference of the SM asymmetry and the prediction according to~\eqref{eq:ASYab} with the current bounds on $C_{4f}$ one can identify three points in energy to fit as much WCs. According to the left panel of Figure~\ref{fig:ASY_pol}, the points are found to be $\sqrt{s_{1,2,3}}=89,93,98~\rm{GeV}$. 
In order to quantitatively estimate the reduction of uncertainty $\Delta C_{4f}$, one can solve the following system for $\alpha=1,2,3$
\begin{equation}
 \label{eq:afb_system}
    \begin{aligned}
\sum_{i\in 4f} \frac{C_i}{\Lambda_\text{NP}^2}\left[\frac{(\sigma_{{\rm{F}}}-\sigma_{{\rm{B}}})_i^{(6)}}{(\sigma_{{\rm{F}}}-\sigma_{{\rm{B}}})_{\rm{SM}}}-\frac{(\sigma_{{\rm{F}}}+\sigma_{{\rm{B}}})_i^{(6)}}{(\sigma_{{\rm{F}}}+\sigma_{{\rm{B}}})_{\rm{SM}}}\right]_\alpha=\frac{\Delta A^{0}_{{\rm{FB}},\alpha}}{A^{0}_{{\rm{FB}},\alpha}}\, ,
\end{aligned}
\end{equation}
assuming the asymmetry to be centred about the SM value with a statistical uncertainty $\Delta A^0_\text{FB}$ estimated with the FCC luminosity $\mathcal{L}_\text{FCC}=\SI{1.4e36}{\cm\tothe{-2}\second\tothe{-1}}$ in six months of run at each $\sqrt{s_\alpha}$. By doing so, we find that the SMEFT shift after the fit is at the level $\delta_\text{SMEFT}\sim 5\times10^{-6}$, enough to reach the precision goal at the $Z$ peak and beyond.\\
If the future colliders physics program will start at energies above $250~\rm{GeV}$, like in the case of ILC, the forward-backward asymmetry is no longer a good observable, as it reaches a constant value. However, such high energy machines are likely to feature polarised $e^\pm$ beams, hence polarisation asymmetries differential in the scattering angle can be exploited. To this end, we build a multivariate Gaussian likelihood whose negative logarithm is distributed as a $\chi^2$
\begin{equation}
    \chi^2 = -2\log L=\sum_{\alpha=1}^n\frac{\left(A_{\rm{pol}}^0-A_{\rm{pol}}^\text{th}(\vec{C}_{4f})\right)_\alpha^2}{(\Delta A_{\rm{pol}}^0)_\alpha^2}\, .
    \label{chi2}
\end{equation} 
 The left-right asymmetry, defined as the difference between the cross section with certain polarisation fractions $\sigma(P_{e^+},P_{e^-})$ and the same flipping the sign of each $P$, is mildly sensitive to $C_{le}$. For this reason, we propose the following asymmetry, obtained by fixing the positron beam polarisation and flipping the electron one
\begin{equation}
     \hspace{-1mm}  A_{\uparrow\downarrow}^-(P_{e^\pm},\cos\theta) = \frac{\dd\sigma(P_{e^+},P_{e^-})-\dd\sigma(P_{e^+},-P_{e^-})}{\dd\sigma(P_{e^+},P_{e^-})+\dd\sigma(P_{e^+},-P_{e^-})} \, .
\end{equation}
Considering the ILC luminosity $\mathcal{L}_{\rm{ILC}}=\SI{1.35e34}{\centi\metre\tothe{-2}\second\tothe{-1}}$ for six months of run, we generate MC replicas of the polarisation asymmetries centred about the SM value with a Gaussian uncertainty given by the statistical error. In the right panel of Figure~\ref{fig:ASY_pol}, we compare the bounds on four-electron coefficients obtained from the region $\chi^2\leq1$, using alternatively $A_{\text{LR},\uparrow\downarrow}$. The up-down asymmetry provides a better constraining power, shrinking the bounds of one order of magnitude more w.r.t. $A_{\text{LR}}$, yielding to $\delta_\text{SMEFT}\sim 10^{-5}$.
\section{Conclusions and outlook}
We have shown that the heavy new physics contribution to the small angle Bhabha scattering as luminosity monitoring process at future $e^+e^-$ colliders is not negligible, given the present knowledge on SMEFT coefficients. In order to reduce the associated theoretical uncertainty, we devise a strategy to better constrain four-electrons coefficients who give the bulk of the contribution, based on asymmetries such that the heavy NP contamination to the luminometry process is reduced below the precision goal. A natural extension of this work is considering the NP contributions to the $e^+e^-\to\gamma\gamma$ process, as well as considering the Muon collider. 

\acknowledgments The authors acknowledge Mauro Chiesa, Guido Montagna, Oreste Nicrosini and Fulvio Piccinini for fruitful collaboration.

\end{document}